\documentclass[twocolumn,aps,prl]{revtex4}
\usepackage{amsmath,graphicx,color,amssymb,bm}

\begin{document}

\title{Molecular  fingerprints in the electronic properties of 
crystalline organic semiconductors: from experiment to theory}

\author{S. Ciuchi$^{1}$, R. C. Hatch$^{3}$, H. H\"ochst$^{4}$, C. Faber$^{2}$,
  X. Blase$^{2}$
 and  S. Fratini$^{2}$} 
\affiliation{$^1$Istituto dei Sistemi Complessi CNR, CNISM and Dipartimento di Fisica,
Universit\`a dell'Aquila, via Vetoio, I-67100 Coppito-L'Aquila, Italy\\
$^2$ Institut N\'eel-CNRS and Universit\'e Joseph Fourier,Bo\^ite
Postale 166, F-38042 Grenoble Cedex 9, France\\
$^3$ Department of Physics and Astronomy, Aarhus University, 8000
Aarhus C, Denmark  \\
$^4$ Synchrotron Radiation Center, University of Wisconsin-Madison, 3731 Schneider Drive, Stoughton, Wisconsin 53589, USA}

\begin{abstract}
By comparing photoemission spectroscopy with a non-perturbative
dynamical mean field theory extension to many-body {\em ab initio}
calculations, we show in the prominent case of
pentacene crystals that an excellent 
agreement with experiment for the
bandwidth, dispersion and lifetime of the hole carrier bands can be
achieved in organic semiconductors 
provided that one properly accounts for the coupling
to molecular vibrational modes and the presence of
disorder. Our findings rationalize the 
growing experimental evidence that even the best
band structure theories based on a many-body treatment of electronic
interactions cannot reproduce the experimental photoemission data
in this important class of materials.
\end{abstract}

\date{\today}

\maketitle

{\it Introduction.--- } Organic semiconductors are key materials
for future applications ranging from flexible electronics to
photovoltaics \cite{Forrest}. 
Because of the weakness of van der Waals  inter-molecular  bonds, 
these materials are commonly believed to stand in an intricate region 
between the molecular limit and the extended band picture,
calling for concepts that go beyond the standard paradigms 
of inorganic semiconductors. 
Experimental 
charge transport and optical studies 
over the past few years 
have indeed given contrasting indications on the nature of the electronic
carriers. 
The existence of low-mass
quasiparticles  indirectly inferred from optical sum rules \cite{Basov}, 
and the apparently ``band-like''  temperature dependent 
electronic mobilities observed in the best crystalline
systems \cite{Podzorov,Sirringhaus,Liu,Minder} 
are difficult to reconcile --- within the framework of band theory
alone --- 
with absolute values of the mobility 
that hardly exceed few tens of $cm^2/Vs$. 
%
The coupling to molecular vibrations as well as the presence
of disorder
are often regarded as the primary causes of
the poor conductive properties of this broad class of
materials \cite{Coropceanu07}, as both
phenomena can slow down the electron motion in the already narrow bands
constructed from $\pi$-intermolecular overlaps. Still, no direct
proof on how such microscopic mechanisms affect the electronic
states has been given to  date. 

As of today, 
while the strongest efforts are devoted to understanding and improving
their 
charge transport characteristics,
a proper description of even
 the most basic electronic properties
such as the band dispersion and carrier lifetime
--- the foundations of our understanding of
conventional semiconductors ---  remains a challenge in  
organic semiconductors.
Recent angle resolved photoemission  (ARPES) 
experiments performed on crystalline samples
\cite{Kakuta,Ohtomo, Hatch,  Hatch09,Machida,Ding} have evidenced that
the best band structure theories 
based on a many-body treatment of electronic interactions 
cannot reproduce the experimental spectra, as they predict
electronic bandwidths that
are  significantly smaller than measured
experimentally.
Here we provide a combined theoretical/experimental analysis
 of the ARPES spectra of pentacene ---
 a material that is often regarded as a model
compound for organic molecular solids. By focusing 
on the full momentum dependence of the spectral features, 
we demonstrate
that the 
interaction of electrons with the internal vibrations
of the molecules
deeply modifies the nature of the extended electronic states.
The multiple vibrational overtones that are commonly
observed in angle-integrated photoemission spectra \cite{Malagoli04,
Yamane,Kera}
translate into a complex fine-structure in momentum space, 
unveiling an unexpectedly large separation of the electronic bands that is
ultimately responsible for the observed increase of the total bandwidth.
Analogous molecular fingerprints are predicted to
affect the electronic bands of broad classes of molecular systems including
organic semiconductors, charge-transfer organic salts and 
doped organic conductors \cite{Yang}.


%

\begin{figure*}
  \centering
\includegraphics[width=18cm]{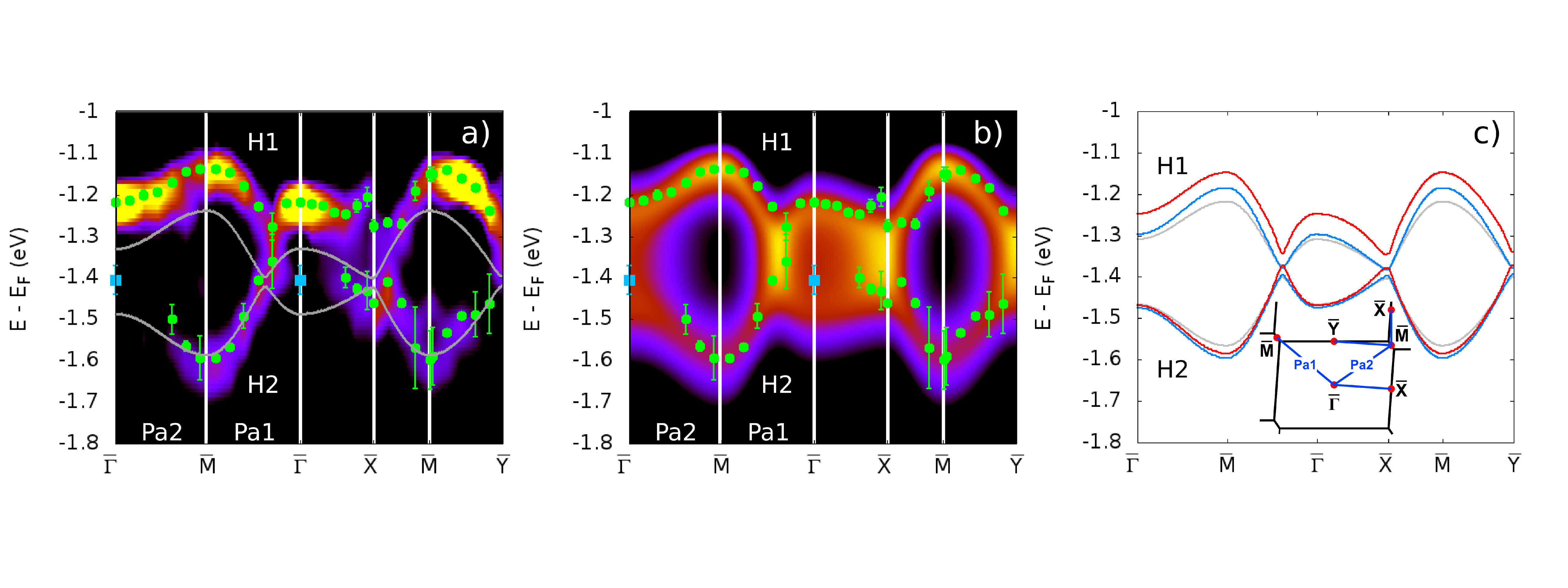}
  \caption{ 
    (a) Color density plot of experimental ARPES data.  The second
derivative of the data has been taken to best visualize
the energy positions of the dispersive features. 
The green dots are the
   peak positions as derived from gaussian fits of the individual
   spectra.
The blue dot 
corresponds to a scan with a different incoming photon
energy, which recovers the  H2 
spectral weight around  $\bar{\Gamma}$ that is hidden 
due to the matrix element dependence on the photon incoming energy
\cite{suppl}.
The  gray line is the {\it ab  initio} band dispersion from
Ref. \onlinecite{Yoshida};
(b)  
    Calculated spectrum in the presence of EMV  interactions and
    disorder (see text).
The dots are the experimental peak
    positions from panel (a); 
(c) Theoretical HOMO band dispersion, as obtained
from the poles of the spectral function.   
{\it Ab initio}
(gray), disordered (blue: $\Delta=75$ meV, $E_P=0$) and with EMV interactions
(red: $\Delta=75$ meV, $E_P=69$ meV).  The inset shows the corresponding path 
in the Brillouin zone.
}
  \label{fig:densplot}
\end{figure*}

{\it Photoemission experiment.---}
In order to probe the effect of electron-molecular vibration interactions
 on the electronic spectra of organic solids, 
ARPES  experiments were performed on in situ grown pentacene films of 
thickness $d>100$\AA \ deposited on Bi(001). This is in contrast with 
previous experiments performed on organic monolayers
\cite{Kakuta,Kera,Yamane}, where the energy-momentum 
dispersion could be seriously affected by the binding to the substrate.
All experiments were carried out at the University of Wisconsin 
Synchrotron Radiation Center (SRC). A detailed
description of the sample preparation process and of the resulting
crystalline film including lattice constants $\mathbf{a}$ and
$\mathbf{b}$ are reported elsewhere \cite{Hatch09,suppl}. 
 The combined photon and electron energy resolution was $\Delta
E\sim 40$ meV and, for $15$ eV photons (used in most experiments) the
electron momentum resolution was $\Delta k_\parallel<0.03$ \AA$^{-1}$.
Fig. \ref{fig:densplot}a shows 
the photoemission spectra 
measured at $T=75$K 
near the top of the highest occupied molecular orbital (HOMO)
as a function of parallel momentum $k_\parallel$ along the path
illustrated in the inset of Fig. \ref{fig:densplot}c. 
Since the lattice structure has two molecules per
unit cell,  
two dispersive branches can be identified, that we denote as H1 and H2.
The dots in Fig. \ref{fig:densplot}a 
are the positions of the two most prominent peaks at each given 
$k_\parallel$, as obtained from gaussian fits to the individual
spectra. 
The solid lines are the band dispersions obtained from
state-of-the-art {\it ab initio} calculations  
for the crystalline film phase which corresponds to the pentacene
polymorph of our sample \cite{Yoshida,shift}.
It is apparent that 
the separation between the  H1 and H2 branches in the experiment 
is much larger than predicted. 
As a consequence the measured total HOMO bandwidth,  
$W_{exp}=450\pm 15$
meV, defined as the distance between the H1 and H2 peaks 
at the $\bar {\mathrm{M}}$ point \cite{suppl}, 
is also much larger than the calculated values  
for the pentacene crystalline film phase, both within DFT  \cite{Yoshida}
($W_0=348$ meV)
and within the 
more accurate GW method \cite{Tiago} ($W_0=360$ meV).  The discrepancy  
is well beyond the experimental uncertainty, which confirms
the systematic findings reported in the most recent ARPES measurements
on pentacene and rubrene  \cite{Kakuta,Ohtomo, Hatch,  Hatch09,Machida,Ding}. 
This indicates that the band-structure calculations alone, even including
electronic many-body correlations as is done in Ref.  
\onlinecite{Tiago}, are not
sufficient to properly describe the observed excitation spectra.

{\it Theoretical modeling.---}
In order to solve the discrepancy between theory and experiment, 
we now 
calculate the photoemission spectral function 
by considering 
the coupling of 
holes in the HOMO band with the
high-frequency, intramolecular vibrations arising from the
carbon-carbon stretching forces.
%
%
We 
use  a non-perturbative, theoretical method 
 in order to avoid any {\it a priori}
assumption on the relative importance of the different phenomena
at work. 
This is a crucial issue since in organic semiconductors the 
characteristic energy scales, namely (a) the electronic bandwith $W_0$,
proportional to the intermolecular electronic transfer energies, (b) the
strength  of the electron-EMV coupling, 
measured by the structural relaxation energy $E_P$ associated with
adding a charge to a molecule,  (c) the molecular vibrational energies 
and (d) the energy fluctuations 
due  to disorder, that cannot be ignored when dealing with real
systems,   are all of comparable magnitude ($\sim 0.1$ eV).
This means that no small parameter can be identified that would allow
application of perturbation theory. 
The microscopic parameters of the EMV interaction specific to  pentacene
are obtained here via
accurate many-body {\it ab initio}  calculations  
within the  GW formalism 
 \cite{GW,Blase11,Faber11b,Lazzeri08,suppl}. This method
has been shown to yield much better electron-phonon coupling 
potentials as compared to  standard DFT-LDA calculations, 
in the case of $\pi$-conjugated fullerenes \cite{Faber11b} and graphene 
\cite{Lazzeri08}. 
We then calculate the resulting photoemission spectrum 
via DMFT  \cite{RMP,depolarone},  which 
is able to address in an unbiased way the
intricate physical regime of interest in organic semiconductors. 
The DMFT scheme also allows one to include the presence of disorder 
that is expected to be relevant in experimental samples,
in the form of spatial fluctuations of the molecular energy levels.
Experimental estimates of the energy spread $\Delta$ 
range from few tens of meV in single crystals
to $\Delta \sim 0.1$ eV or above in amorphous samples  \cite{Kalb}. Our
crystalline films should be located between these two limits 
as indicated by the sharpness of our  reflection high energy electron
diffraction (RHEED) patterns.
Finally, the interaction with low-frequency vibrations 
of the molecules
 \cite{Malagoli04,Girlando11}, 
 in the range $\hbar \omega \lesssim 40$ meV,  
is   effectively   included in the calculation via an increased  
value of $\Delta$.
These low-frequency modes related both to the large molecular size and to the
mechanical softness of the material
do not give rise to observable features in the
photoemission spectra, as their energy lies below the experimental resolution,
but they do act as an additional intrinsic source of fluctuation for the
molecular energy levels  \cite{Ciuchi11,HoHu}.

{\it Molecular origin of the band separation.---}
Fig. \ref{fig:densplot}b is a color density plot of the hole spectral
function obtained from the theory including both molecular and
itinerant aspects on the same footing, as described in the preceding
paragraph.  
We take  the calculated value
$E_P=69$ meV for the molecular relaxation energy (corresponding to 
a reorganization energy $2E_P=138$ meV), $\hbar\Omega=174$ meV
for the average  energy of intra-molecular vibrations \cite{suppl}, a
non-interacting bandwidth $W_0=348$ meV  
 and assume a total  disorder strength  $\Delta= 75$ meV.
The similarity with the experimental spectra is remarkable. 
The most striking point is that the 
separation between the H1 and H2 bands 
is much larger than in the
band-structure prediction, restoring 
the agreement with the locus of the experimental peaks (dots).  
Also in agreement with the experiment,  
the features of the calculated spectrum exhibit a 
considerable broadening, 
a point that will be discussed below.

To ascertain the microscopic origin of the observed H1/H2 separation,
Fig. \ref{fig:densplot}c shows the HOMO band dispersion 
obtained from the theory
upon the subsequent inclusion of
 disorder (blue curve) and EMV interactions  (red curve) on top of the
 band structure as parametrized in Ref. \onlinecite{Yoshida}.
Disorder in the molecular energy levels causes a rather uniform 
increase of the band dispersion, but does not modify the
overall band shape  \cite{CPA}.
Instead, adding the interaction with high-frequency molecular vibrations
causes a sizable upwards shift of the H1 band, 
leaving the H2 band dispersion  essentially unchanged from 
the non-interacting case.  
This shift is caused by the molecular relaxation energy gain $E_P$,
that is reflected in the extended states of the solid 
through a  stabilization of the H1 branch.
We conclude that the interplay between EMV coupling and disorder 
is  at the origin of 
the large H1/H2 separation observed in the experiment. The present analysis
 confirms the results obtained in Ref. \cite{Ciuchi11} based on a related
 one-dimensional model.

\begin{figure}
  \centering
  \includegraphics[angle=0,width=6cm]{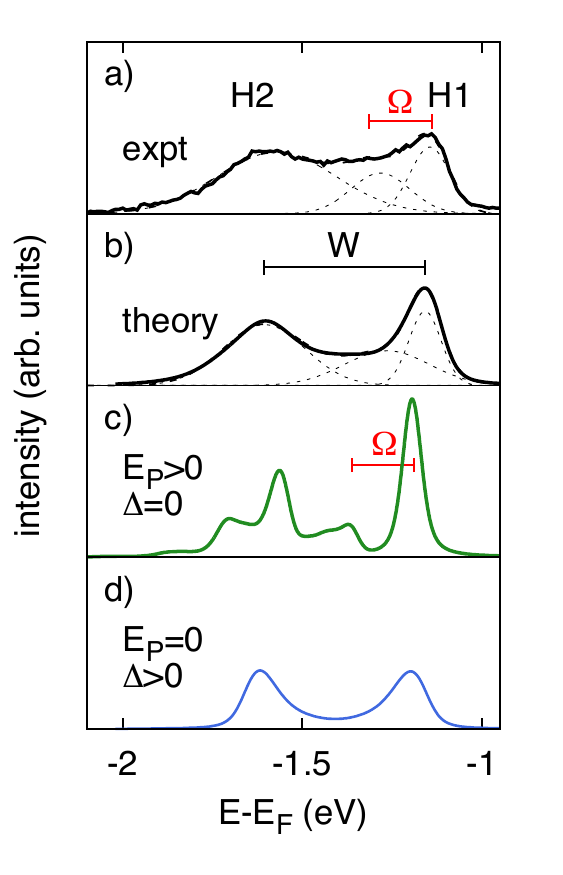}  
  \caption{ \label{fig:comparison} 
(a) Photoemission spectrum measured at the $\bar{\mathrm{M}}$ point. The dashed
line is a fit of the data with three gaussian peaks. (b)
Calculated spectrum and fit. 
 Parameters are $E_P=69$ meV and
    $\Delta=75$ meV. $W$ is the total bandwidth defined as the distance 
between the H1 and H2 peak. 
 (c) The fine structure
induced by the EMV interaction is unveiled by removing the disorder in
the calculation.
(d) The calculated
spectrum in the absence of EMV interactions, 
but with the same degree of disorder. In (b,c,d) 
the spectrum has been convoluted with a Gaussian of FWHM
$=40$ meV to mimic the experimental resolution.
}
\end{figure}

{\it Spectral hallmarks of EMV interactions.---}
Having shown that the main dispersive 
features of the ARPES data
can be explained by the proper inclusion of
interactions with high-frequency molecular vibrations and a moderate amount 
of disorder, we now proceed to show that additional 
hallmarks of the EMV
coupling are  seen in the angle-resolved spectra, providing
support to
the proposed scenario. 
Fig. \ref{fig:comparison}a reports the photoemission intensity
measured at the $\bar{\mathrm{M}}$ point (full line), and Fig. \ref{fig:comparison}b
is the corresponding calculated spectrum. 
In addition to the H1 and H2 peaks that are clearly resolved, 
an extra feature can be actually recognized, whose existence  
is substantiated by fitting the spectra with three gaussian peaks 
(dashed lines). Being located at a distance $\simeq \hbar
\Omega$ from the main H1 peak, 
it is tempting to associate this feature with a vibrational overtone 
of the main electronic excitation.
That this is indeed the case is demonstrated 
by repeating the theoretical calculation in the absence of disorder, 
 which reveals all the fine structure that
is otherwise  smeared out by the energy fluctuations 
(Fig. \ref{fig:comparison}c).  The overtone actually
disappears as expected when the EMV interaction is turned off
(Fig. \ref{fig:comparison}d).


Our analysis shows that the 
presence of multiple vibrational overtones is also at the origin
of the large broadening of the H2 peaks  observed in the experiment, 
whose tails extend to very large binding energies
(see both Fig.  \ref{fig:comparison}a and \ref{fig:comparison}b). 
The reduced lifetime  of the H2 states
again results from a non-trivial interplay between 
vibrational shakeoff excitations  and disorder.
The individual resonances contributing to the H2 weight are not  
resolved in the experiment, as these 
merge into a single broad feature due to the 
presence of disorder. Such vibrational overtones are however
clearly revealed in the calculation on a 
perfectly ordered crystal (Fig.  \ref{fig:comparison}c) and could possibly be 
detected in future experiments on cleaner samples.  A more detailed
discussion on the interplay between EMV interactions and disorder,
not limited to the representative $\bar{\mathrm{M}}$ point considered here, 
is presented in \cite{suppl}.

\begin{figure}
  \centering
\includegraphics[angle=270,width=9cm]{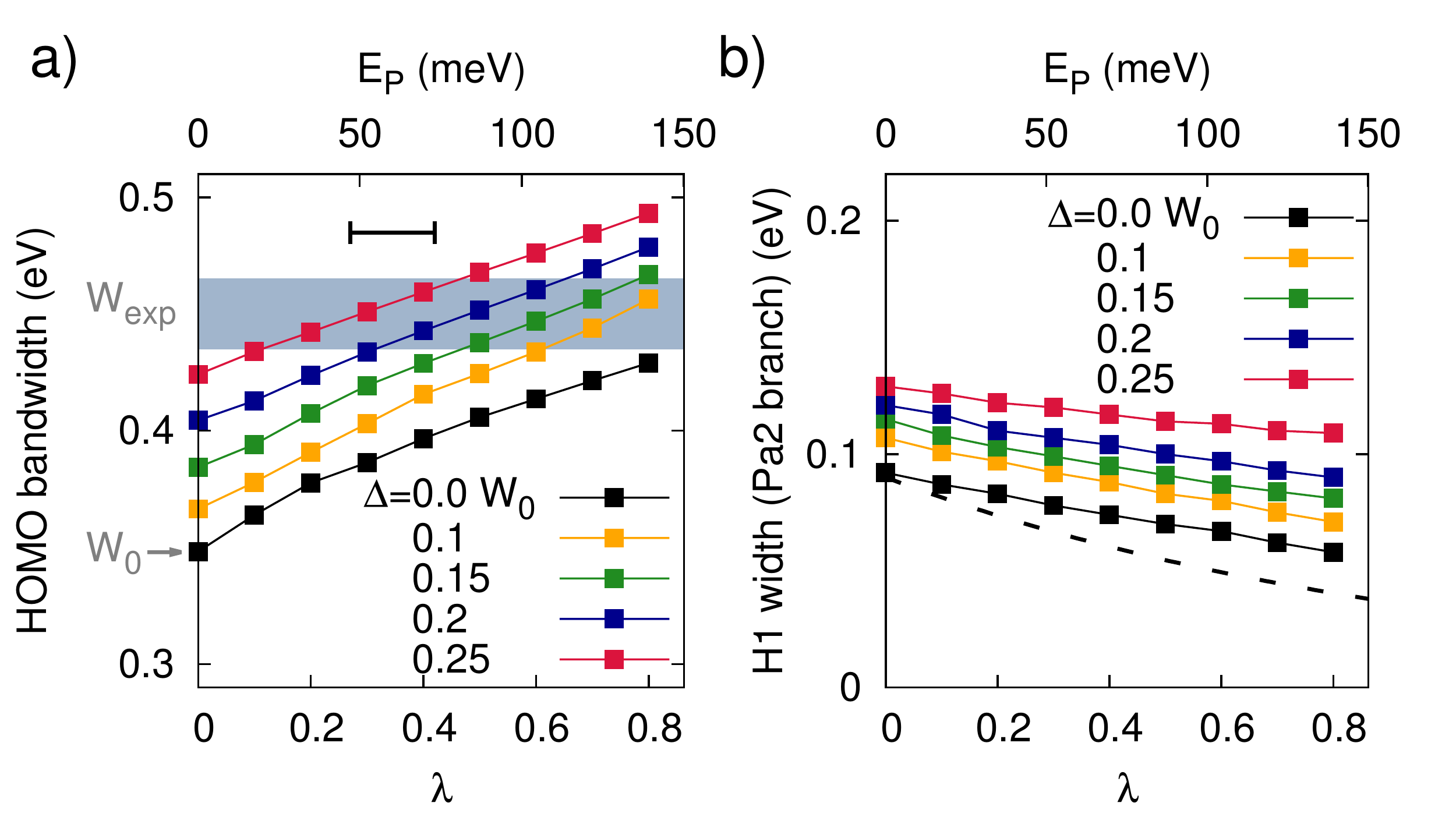}  
  \caption{\label{fig:bands} 
(a) Calculated  HOMO bandwidth  
as a function of the EMV coupling strength
for different values of the disorder parameter. 
 The shaded area is the measured value (see text). 
The horizontal bar indicates the range of calculated $E_P$ values in
the literature. 
(b) Renormalization of the H1 band along Pa2
($\bar{\mathrm{M}}-\bar{\Gamma}$ distance, see Fig. \ref{fig:densplot}).
The dashed line is the polaronic renormalization
expected in the molecular limit.
 }
\end{figure}

{\it Quantitative analysis and discussion.---}
Finally we show that combining  model
Hamiltonian studies with {\it ab initio} calculations as we have done here
can actually be used to 
extract quantitative information on the relevant microscopic
parameters of the materials directly from the experiment.
Fig. \ref{fig:bands}a illustrates the
evolution of the total calculated HOMO 
bandwidth $W$ resulting from the molecular relaxation phenomenon
as a function of the EMV coupling strength 
$\lambda=2E_P/W_0$, 
for different values of the disorder parameter $\Delta$. 
%
Taking the
calculated value, $E_P=69$ meV, our experimental data are
compatible with a total disorder strength $\Delta = 75 \pm 15$
meV, in agreement with the values available in the literature
 \cite{Kalb}.
 As mentioned previously, this is an effective parameter that also
includes the intrinsic fluctuations originating
from the low-frequency molecular vibrations,
whose contribution can be estimated to $\approx 20$ meV 
 \cite{Girlando11,Ciuchi11}. The  estimate $\Delta=75$ meV given above 
therefore constitutes an
upper bound to the actual structural disorder present in the sample.    
We note that the value of the molecular relaxation energy itself could be
extracted  from the experiment if a precise independent estimate of 
the disorder fluctuations were available.

Beyond these quantitative considerations, 
the present study  clearly establishes the theoretical framework that
should be worked with in   
upcoming studies of transport properties, since a good description of
the momentum and energy  
dependence of carriers is a necessary basis to properly assess
mobility properties.
In particular, 
our results show unequivocally that the total HOMO bandwidth {\it
  increases} with the EMV coupling. This behavior  is opposite 
to the  bandwidth {\it decrease} that constitutes the basis of most
theories of charge transport in organic  semiconductors, but that   
is only expected to apply in the molecular limit, $W_0 \ll \hbar \Omega$
 \cite{HolsteinII,Mahan,Coropceanu07}.
Upon closer inspection, we find that a 
partial band shrinking occurs in a restricted energy range,
roughly within a window
$\hbar\Omega$ from the top of the band (the ground state for
holes) \cite{depolarone}. 
The width of the H1 branch along  Pa2, reported in Fig. \ref{fig:bands}b,
indeed shows a moderate decrease upon increasing the EMV coupling,
which however does not obey the textbook \cite{Mahan} 
exponential renormalization 
$\propto \exp (-E_P/\hbar\Omega)$ (dashed line). 
 The fact that  the phenomenon of band narrowing  
is shown here to play a reduced role 
should be taken into
account when constructing a theory for charge transport in
organic semiconductors.



{\it Concluding remarks.---}
There is growing experimental evidence that the
electronic properties of  organic molecular semiconductors  
cannot be understood within the commonly accepted band structure
picture that prevails for inorganic semiconductors. 
This discrepancy 
entails that a consistent
description of the extended states in molecular solids 
should take full account of the microscopic processes taking place at the
molecular level. 
The analysis presented here indeed
demonstrates 
that the proper inclusion of 
the interaction with molecular vibrations 
and disorder, beyond 
electronic band theory calculations,
provides a remarkably accurate 
description of the experimental photoemission spectra in pentacene.
The coexistence of dispersive features characteristic 
of the extended band regime, together with 
 multi-phonon shakeoff resonances reminiscent 
of the molecular spectra,
 provides solid spectroscopic evidence for
the widespread idea that organic materials 
are located in a crossover region where 
the band and molecular characters are inextricably linked.

{\it Acknowledgments.---} {\it Ab initio} calculations have been performed on the supercomputing IDRIS
and CIMENT facilities. This work is based in part upon research
conducted at the Synchrotron Radiation Center,  
University of Wisconsin-Madison, which is supported by the National
Science Foundation under Award No. DMR-0537588. We acknowledge
stimulating discussions with D.L. Huber as well as financial support
by the Lundbeck foundation.  We thank A. Morpurgo and G. Profeta 
for a critical reading  of the manuscript.

\bigskip


\begin{thebibliography}{30}
\bibitem{CPA}This is  a known result for disordered systems that 
can be formally understood, for example,   
by solving for the electron propagator in the
coherent potential approximation (CPA), see Ref. \cite{HoHu}
\end{thebibliography}
\end{document}